\documentclass[journal]{IEEEtran}
\ifCLASSINFOpdf
\else
   \usepackage[dvips]{graphicx}
\fi
\usepackage{url}

\usepackage{graphicx}

\usepackage{algorithm}
\usepackage{algorithmic}
\usepackage{mathtools}
\usepackage{booktabs}
\usepackage{cite}

\usepackage{amssymb}

\usepackage{graphicx}    % For including images
\usepackage{subcaption}  % For subfigures
\usepackage{enumitem}
\usepackage{amsmath}

\begin{document}

\title{SegINR: Segment-wise Implicit Neural Representation for Sequence Alignment in Neural Text-to-Speech}

\author{Minchan Kim~\IEEEmembership{{Student Member,~IEEE,}} Myeonghun Jeong~\IEEEmembership{{Student Member,~IEEE,}} Joun Yeop Lee, and Nam Soo Kim~\IEEEmembership{Senior Member,~IEEE}

\thanks{This work was supported by Samsung Research, Samsung Electronics Co.,Ltd.}
\thanks{Minchan Kim, Myeonghun Jeong, and Nam Soo Kim are with the Department of Electrical and Computer Engineering and with the Institute of New Media and Communications, Seoul National University, Seoul 08826, South Korea (e-mail: mckim@hi.snu.ac.kr; mhjeong@hi.snu.ac.kr; nkim@snu.ac.kr)}
\thanks{Joun Yeop Lee is with Samsung Research, Seoul, 06765, Republic of Korea (e-mail: jounyeop.lee@samsung.com)
}}

\markboth{Journal of \LaTeX\ Class Files, Vol. 14, No. 8, August 2015}
{Shell \MakeLowercase{\textit{et al.}}: Bare Demo of IEEEtran.cls for IEEE Journals}
\maketitle

\begin{abstract}
We present SegINR, a novel approach to neural Text-to-Speech (TTS) that addresses sequence alignment without relying on an auxiliary duration predictor and complex autoregressive (AR) or non-autoregressive (NAR) frame-level sequence modeling. SegINR simplifies the process by converting text sequences directly into frame-level features. It leverages an optimal text encoder to extract embeddings, transforming each into a segment of frame-level features using a conditional implicit neural representation (INR). This method, named segment-wise INR~(SegINR), models temporal dynamics within each segment and autonomously defines segment boundaries, reducing computational costs. We integrate SegINR into a two-stage TTS framework, using it for semantic token prediction. Our experiments in zero-shot adaptive TTS scenarios demonstrate that SegINR outperforms conventional methods in speech quality with computational efficiency.
\end{abstract}


\IEEEpeerreviewmaketitle

\section{Introduction}
\IEEEPARstart{N}{eural} Text-to-Speech (TTS) models inherently address the alignment problem by regulating sequence length, expanding text length into speech length based on the irregular monotonic alignment between text and speech. This alignment problem is typically tackled using intermediate frame-level features (e.g., mel-spectrogram, semantic tokens~\cite{kharitonov2023speak, lee2024high}, acoustic tokens\cite{defossez2023high, yang2023hifi}) rather than raw waveforms. Conventional TTS models can be categorized into two types depending on alignment modeling: autoregressive (AR) and duration-based non-autoregressive (NAR) methods. AR models extend frames sequentially, dynamically determining the relevant parts of the text features, including attention-based sequence-to-sequence (seq2seq) models~\cite{shen2018natural, kharitonov2023speak, wang2023neural} and transducers~\cite{kim2023transduce, chen2021speech, du2024vall}. However, AR models have drawbacks such as requiring recurrency during inference, which leads to slow inference and error propagation, especially with misalignment~\cite{valentini2021detection}. In contrast, duration-based NAR models~\cite{ren2020fastspeech, kim2021conditional, popov2021grad} utilize explicit phoneme durations for length regulation, expanding text embedding sequences to align with frame-level features based on duration, then converting them into frame-level features in parallel using various generative models. These models use ground truth alignment acquired from forced alignment algorithms~\cite{mcauliffe2017montreal, kim2020glow, badlani2022one} during training and predicted duration obtained from an auxiliary duration predictor during inference, which requires additional duration modeling and induces cascading errors associated with the duration predictor.

In this paper, we propose a novel method that converts text sequences into frame-level features without the need for both an auxiliary duration predictor and complex frame-level sequence modeling. We assume that an optimal text encoder can extract a text embedding sequence where each frame contains sufficient information for its assigned segments of frame-level features. Following this assumption, we decompose the seq2seq task into a set of embedding-to-segment (emb2seg) conversions, which transform a text embedding into a segment of frame-level features. We build each emb2seg conversion model based on implicit neural representation~(INR)~\cite{sitzmann2020implicit, mildenhall2021nerf, park2019deepsdf, pmlr-v162-dupont22a}. INR is a multi-layer perceptron (MLP) model that represents continuous signals as a function of coordinates. We construct a conditional INR that takes the time index $i$ within the segment as input and returns the $i_{th}$ frame of the segment, using the text embedding as a conditioning factor. This conditional INR, named Segment-wise INR (SegINR), represents the temporal dynamics of the frame-level feature within a segment assigned to each text unit. SegINR replaces length-expanded sequence modeling with building a function space of time. Additionally, we introduce an end of segment token $\varnothing$, allowing INR to automatically determine its own duration. By jointly predicting the output sequence and the $\varnothing$ token, the model can define segment boundaries autonomously without using an external duration predictor. SegINR significantly reduces the computational cost of length-extended sequence modeling, as the proposed method only requires text-level sequence modeling. Since the SegINR consists of shallow MLP layers without receptive field, the computational requirements are minimal compared to sequence modeling at the frame-level feature length. The final output sequence is a concatenation of all segments generated independently by the SegINR.

We explore the application of SegINR within a two-stage TTS framework~\cite{lee2024high}. As sequence expansion based on duration is conducted in the semantic token prediction stage, we adopt SegINR for semantic token prediction from text, replacing the token transducer. Using semantic tokens as target features, which contain linguistic and coarse-grained speech information, helps alleviate the inevitable discontinuity at segment boundaries. The discrete output space also facilitates the construction of a joint space for the output and the $\varnothing$ token. We then generate waveforms from semantic tokens using a masked language model~\cite{jeong2024efficient}. Our experiments investigated the proposed model in zero-shot adaptive TTS scenario. We demonstrate that the proposed method surpasses other conventional methods. Our generated samples are available on the demo page\footnote{https://gannnn123.github.io/seginr}.

\section{Backgrounds}
\subsection{Implicit Neural Representation~(INR)}
Implicit Neural Representations (INRs) are neural networks that parameterize a field in continuous coordinates. They provide a way to represent complex, high-dimensional data with a small number of learnable parameters, which can be used for various tasks. For instance, a colored 2D image can be represented as $R^2 \rightarrow R^3$, where pixel coordinates are mapped to RGB intensity using neural networks. INRs have been widely used for various purposes, including data compression~\cite{dupont2021coin,dupont2022coin++}, 3D rendering~\cite{mildenhall2021nerf, park2019deepsdf, jun2023shap}, and generative models~\cite{pmlr-v162-dupont22a, bauer2023spatial, jun2023shap}, with ongoing developments in representing fine-grained details and high-order derivatives~\cite{sitzmann2020implicit, tancik2020fourier}.

We focus on the concept of conditional INRs, which represent data with a conditioning embedding and shared parameters of the INR. In conditional INRs, general prior knowledge is embedded in the parameters of the INR, while specific characteristics of individual examples are encoded in the conditioning embedding. %We exploit the basic idea and structure of conditional INRs to solve the embedding-to-segment (emb2seg) transformation problem.

\subsection{Length Regulation in TTS}
%In this section, we introduce the characteristics of commonly used methods for alignment modeling within TTS.

\subsubsection{Attention-based AR Models~\cite{shen2018natural,li2019neural,kharitonov2023speak,wang2023neural}}
Attention-based AR models calculate alignment using an attention mechanism~\cite{bahdanau2014neural, soydaner2022attention}. Instead of defining explicit alignment and duration, they rely on attention scores. These models do not require external forced alignment algorithms or duration predictors. However, the autoregressive nature of these models induces slow inference. Moreover, the attention mechanism cannot guarantee monotonic constraints, which can result in alignment failures~\cite{valentini2021detection}.

\subsubsection{Transducer~\cite{kim2023transduce, chen2021speech, du2024vall}}
Transducers are designed for seq2seq tasks with monotonic alignment, such as speech recognition~\cite{graves2012sequence}. Previous works have explored the use of transducers in TTS~\cite{kim2023transduce, chen2021speech, du2024vall}. Transducers build an alignment lattice and formulate the conditional likelihood as the marginalization of all possible paths. They use a special blank token $\varnothing$ for transitions to the next input frame, which inspired our proposed method. Similar to attention-based models, transducers operate in an AR manner, resulting in slow inference.

\subsubsection{Duration-based NAR Models~\cite{ren2020fastspeech, kim2021conditional, popov2021grad}}
Duration-based NAR models define explicit durations for text units. Text embeddings are duplicated according to their durations and then decoded to generate the output sequence. These models have the advantage of parallel generation with fast inference speed. However, they require external duration information obtained from forced alignment algorithms~\cite{mcauliffe2017montreal, kim2020glow, badlani2022one} during training and duration predictors during inference. This adds an additional task of building a duration predictor, and the mismatch between training and inference can lead to degradation due to cascading errors.

\section{Method}
\begin{figure*}[h!]
  \centering
  \includegraphics[width=\textwidth]{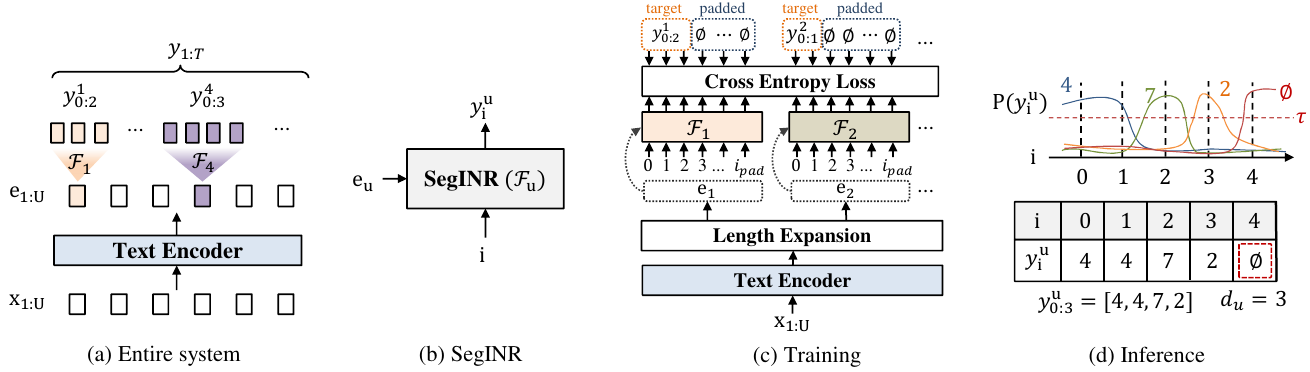}
  \caption{Illustration of SegINR and its application for semantic token prediction: (a) overall concept of SegINR, (b) structure of SegINR, (c) training method for semantic token prediction, (d) inference method for semantic token prediction.}
  \label{fig1}
\end{figure*}

\subsection{Segment-wise Implicit Neural Representation~(SegINR)}

Given an input text sequence $\textbf{x}_{1:U}$ and the corresponding frame-level features $\textbf{y}_{1:T}$, our objective is to construct a model that converts $\textbf{x}_{1:U}$ into $\textbf{y}_{1:T}$. Adhering to the monotonic alignment constraint between text and speech, we define the duration as $\textbf{d}_{1:U}\subseteq \mathbb{Z}_{\geq 0}$, where $\sum_{u=1}^{U} \textbf{d}_u = T$. Each input text token $\textbf{x}_u$ is aligned to the segmented sequence $\textbf{y}_{0:d_u}^{u}$, and as a result, $\textbf{y}_{1:T}$ is formulated as a concatenation of all segments, 
\begin{equation} \label{eq:loss1}
\textbf{y}_{1:T} = \textbf{y}_{0:d_1}^{1} \mid \textbf{y}_{0:d_2}^{2} \mid \ldots \mid \textbf{y}_{0:d_U}^{U}.
\end{equation}
Regarding the text embedding sequence $\textbf{e}_{1:U}$ obtained from a text encoder, we assume that each $\textbf{e}_u$ can contain sufficient information for generating $\textbf{y}_{0:d_u}^{u}$. Then, we breakdown the seq2seq problem into a set of emb2seg problems: generating $\textbf{y}_{0:d_u}^{u}$ from $\textbf{e}_u$, which also includes determining $\textbf{d}_u$.

We address the emb2seg problem using a conditional Implicit Neural Representation (INR) named SegINR. SegINR represents the function of time index $i$: $\mathcal{F}_u(i; \textbf{e}_u, \theta) = \textbf{y}_{i}^{u}$, where $i \in \mathbb{R}$ such that $0 \leq i \leq d_u$. $\mathcal{F}_u$ leverages the inherent continuity of sequences to efficiently represent the temporal dynamics of each segment. Notably, the time index $i$ is a real-valued scalar, even though we use integer values in our framework. Inspired by the transducer model, we also allow $\mathcal{F}_u$ to predict a special token $\varnothing$, indicating the end of a segment, to automatically determine the domain $[0, \textbf{d}_u]$. Thus, we construct the joint output space $\mathcal{Y} \cup \{\varnothing\}$ and set $\mathcal{F}_u(\textbf{d}_u + 1; \textbf{e}_u, \theta) = \varnothing$. Consequently, $\textbf{d}_u$ is determined as the largest index before predicting $\varnothing$. After generating all segments using $\mathcal{F}_u$ independently from $\textbf{e}_{1:U}$, the entire sequence is constructed by concatenating the generated segments as in (1). The proposed framework is illustrated in Fig.~\ref{fig1}. (a) and (b).

\subsection{Application}
\subsubsection{Semantic Token Prediction}
We integrate SegINR within a two-stage TTS framework~\cite{lee2024high}, which includes text-to-semantic token and semantic-to-acoustic token stages. This framework separates coarse-grained linguistic modeling with alignment from fine-grained acoustic modeling, simplifying the modeling process and enhancing performance. Since alignment modeling and sequence expansion are handled in the first stage, we employ SegINR for semantic token prediction. Using semantic tokens as target frame-level features, which contain linguistic and coarse-grained speech information, helps mitigate the inevitable discontinuity at segment boundaries. The discrete output space also facilitates the construction of a joint space for the output and the $\varnothing$ token. Consequently, the output space of the proposed SegINR consists of $|\mathcal{Y}|+1$ classes of categorical distribution, where $\mathcal{Y}$ represents the set of semantic tokens. We then generate waveforms from the semantic tokens using a masked language model~\cite{jeong2024efficient}.

\subsubsection{Architecture}
The entire semantic token prediction model consists of a text encoder and SegINR. The text encoder is built using conformer blocks~\cite{gulati2020conformer}, while SegINR utilizes a modulated SIREN structure inspired by Coin++~\cite{dupont2022coin++}. The modulated SIREN comprises MLP layers with sine activation, modulated by a conditioning embedding $\textbf{e}_u$.

\subsubsection{Training}
We jointly train the text encoder and SegINR using single training loss. To calculate the ground truth duration, we utilize the Token Transducer++~\cite{lee2024high}, a transducer designed for text-to-semantic token translation. Specifically, we calculate the most probable path in the alignment lattice of the transducer using the Viterbi algorithm, summing up the number of frames assigned to each phoneme.

We not only enforce the condition $\mathcal{F}_u(\textbf{d}_u + 1; \textbf{e}_u, \theta) = \varnothing$ but also perform auxiliary training with the condition $\mathcal{F}_u(i; \textbf{e}_u, \theta) = \varnothing$ for $\textbf{d}_u + 1 < i < i_{pad}$, where $i_{pad}$ is a constant representing a sufficiently large padding number for $\varnothing$. This auxiliary training helps to ensure the consistent output of $\varnothing$ for indices greater than $\textbf{d}_u + 1$, thereby improving the stability of the inference process.

Although SegINR is not a sequence model but an MLP model, we train it in a pseudo sequential manner, which is actually implemented at the batch level for convenience, as illustrated in Fig.~\ref{fig1}(c). This approach is similar to that used in non-autoregressive~(NAR) TTS models. After extracting text embeddings from the text encoder, we expand each text embedding $\textbf{e}_u$ for $i_{pad}$ times and create a pseudo time index sequence. We then feed both pseudo input sequences and the expanded text embeddings, then a pseudo output sequence is returned. The entire model is trained using cross-entropy loss between the pseudo output sequence and the target sequence, which is the concatenation of all $\varnothing$-padded semantic token segments.

\subsubsection{Inference}
Once we obtain $\textbf{e}_{1:U}$ from the text encoder, we generate $\textbf{y}_{1:T}$ using SegINR, as illustrated in Fig.~\ref{fig1}~(d). During inference, SegINR returns the most probable semantic token at each time step if the estimated probability of $\varnothing$ is below a threshold $\tau$; otherwise, it returns $\varnothing$. A key advantage of SegINR is that it fits well with both the streaming and parallel inference frameworks. 1) In a streaming scenario, we sequentially decode $\mathcal{F}_u$ by increasing $i$ until $\varnothing$ is returned. Then we move on to next $\mathcal{F}_{u+1}$ until reaching $U$. This process resembles the inference phase of transducers, but operates without recurrency. 2) For parallel decoding, we define $i_{\text{max}}$, the maximum duration per text unit, for every $\mathcal{F}_u$. We then generate all outputs in parallel by injecting $[0, 1, 2, \dots, i_{\text{max}}]$ for all $\mathcal{F}_u$ in a batch process. We take only the valid outputs, stopping at the first $\varnothing$. If $\varnothing$ does not appear until $i_{\text{max}}$, we set $\textbf{d}_u=i_{\text{max}}$. Although this approach involves wasted computation of $\sum_{u=1}^{U} (i_{\text{max}}-\textbf{d}_u)$, SegINR's low computational requirements result in significantly faster inference compared to other sequence-level decoding methods.

\section{Experiments}
\subsection{Experimental Setting}
We conducted experiments on zero-shot adaptive TTS, following the experimental settings of previous work~\cite{lee2024high}. We used the same semantic tokens; the indices of $k$-means clustering on the wav2vec2.0-XLSR model~\cite{conneau2020unsupervised} with $k=512$. We built the semantic token prediction model using SegINR, replacing the Token Transducer++ in \cite{lee2024high}, and employed G-MLM~\cite{jeong2024efficient} for semantic-to-acoustic token conversion. The proposed model was trained on all training subsets of the LibriTTS corpus and evaluated on test subsets.

\subsubsection{Implementation Details}
For semantic token prediction, we used the same text encoder structure of the Token Transducer++ which consist of a conformer blocks~\cite{gulati2020conformer}. The dimension of the text embedding $\textbf{e}_u$ is 384. For SegINR, we implemented a modulated SIREN with three layers of MLP, each with a hidden dimension of 256 using the official code from Coin++~\cite{dupont2022coin++}\footnote{\url{https://github.com/EmilienDupont/coinpp}}. Through our experiments, we found that the SIREN activation frequency of $w_0 = 1.0$ performed well. For zero-shot adaptation, we added a reference encoder with the same structure as described in \cite{lee2024high}. During training, the reference encoder processes a randomly cropped 3-second segment of the target speech as in \cite{lee2024high}. The resulting reference embedding is then globally added at the beginning of the text encoder to condition prosody information. Also, we set $i_{pad}=20$ for training SegINR, and $i_{max}=20$ and $\tau=0.5$ for parallel inference. We trained the proposed model for 50 epochs with dynamic batch size containing up to 240 seconds.

\subsubsection{Baselines}
We used three baseline models for performance comparison: VITS~\cite{kim2021conditional} representing an NAR model, VALLE-X~\cite{zhang2023speak} representing an AR model, and the model proposed by Lee et al.~\cite{lee2024high}. To adapt the baseline VITS to the zero-shot scenario, we incorporated the same reference encoder structure as in our proposed model. For VALLE-X, we used the open-source implementation\footnote{\url{https://github.com/Plachtaa/VALL-E-X}}. The model by Lee et al.~\cite{lee2024high} served as the primary baseline in our work, as we shared the same semantic-to-acoustic token stage but differing in the semantic token prediction models: the Token Transducer++.

\subsection{Results: Zero-Shot Adaptive TTS}
We evaluated the mean opinion score~(MOS), similarity MOS~(SMOS), character error rate~(CER), speaker embedding cosine similarity~(SECS), and real-time factor~(RTF). The MOS test measured perceptual speech quality, rated on a scale from 1 to 5 by 14 testers. In the SMOS test, the same testers rated the similarity in speaker characteristics and prosody between the reference speech and the synthesized samples. The CER assessed the intelligibility of the generated speech, which we calculated using the Whisper large model~\cite{radford2023robust}. For SECS, we used the pre-trained WavLM large speaker verification model\footnote{\url{https://github.com/microsoft/UniSpeech/tree/main/downstreams/speaker_verification}} to calculate speaker similarity between the generated speech and the reference speech. For both objective evaluations, we randomly selected 800 pairs of text and reference speech from the test set.

The results are shown in Table~\ref{result}. The proposed model outperformed baseline models across most metrics. VALLE-X had the highest CER, primarily due to misalignment issues common in attention-based AR models. Since SECS is more influenced by the semantic-to-acoustic token conversion rather than semantic token generation, the model proposed by Lee et al.~\cite{lee2024high} achieved nearly identical scores to our proposed model, as they share the same post-processing models. In terms of inference speed, VITS demonstrated the highest RTF. However, when comparing only the semantic token prediction part between the model by Lee et al.\cite{lee2024high} and the proposed SegINR, the Token Transducer++ in \cite{lee2024high} showed an RTF of 22.35, while SegINR achieved \textbf{134.85}. This substantial difference indicates the feasibility and computational efficiency of SegINR for sequence alignment, even though the other part, G-MLM, occupies a significant portion in our current framework.

\begin{table}
\caption{Results of zero-shot adaptive TTS. MOS and SMOS are represented with 95\% confidence intervals. RTF is calculated by a Quadro RTX8000 GPU.}
\label{result}
\centering
% \begin{tabular}{@{}lc@{}c@{}rc@{}c@{}}
%\begin{tabular}{l r r r r}
\begin{tabular}{l c c c c c}
\toprule
\textbf{Method} & \textbf{MOS} & \textbf{SMOS} & \textbf{CER}(\%)& \textbf{SECS}& \textbf{RTF}\\ 
\midrule
Ground Truth & 4.51\footnotesize{$\pm$0.07} & 4.38\footnotesize{$\pm$0.08} & 1.56 & 0.678 & - \\
\midrule
VITS     & 3.60\footnotesize{$\pm$0.07} & 4.04\footnotesize{$\pm$0.08} & 5.80 & 0.375 & \textbf{70.12}   \\
VALLE-X    & 3.71\footnotesize{$\pm$0.09} & 4.17\footnotesize{$\pm$0.08}  & 10.58 & 0.426 & 1.89 \\
Lee et al.~\cite{lee2024high}     & 4.11\footnotesize{$\pm$0.09} & 4.44\footnotesize{$\pm$0.07}  & 3.55 & \textbf{0.463} & 6.76\\
\midrule
Proposed  & \textbf{4.27}\footnotesize{\textbf{$\pm$0.08}} & \textbf{4.51}\footnotesize{\textbf{$\pm$0.07}} & \textbf{3.14} & 0.461 & 8.72  \\
\bottomrule
\end{tabular}
\end{table}

\subsection{Ablation: Training and Inference Schemes}
\begin{figure}[h!]
    \centering
    % (a)
    \begin{subfigure}[b]{0.24\textwidth}
        \centering
        \includegraphics[width=\textwidth]{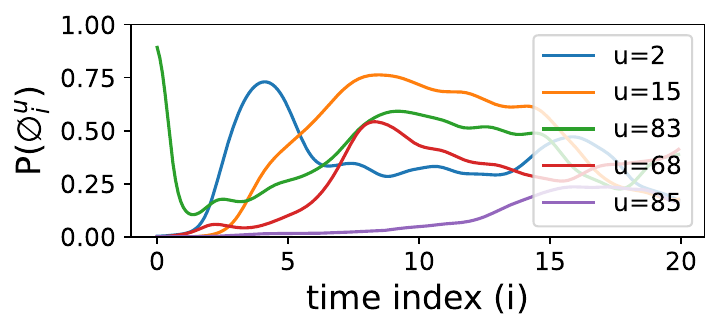}
        \captionsetup{skip=-2pt}
        \caption{P$(\varnothing_i^u)$: w/o padded training}
        \label{fig:a}
    \end{subfigure}
    % (b)
    \begin{subfigure}[b]{0.24\textwidth}
        \centering
        \includegraphics[width=\textwidth]{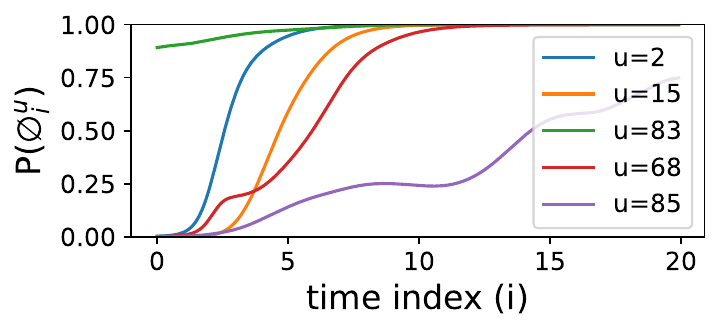}
        \captionsetup{skip=-2pt}
        \caption{P$(\varnothing_i^u)$: with padded training}
        \label{fig:b}
    \end{subfigure}
    
    \vspace{0.2cm} % Vertical space between rows
    
    % (c)
    \begin{subfigure}[b]{0.24\textwidth}
        \centering
        \includegraphics[width=\textwidth]{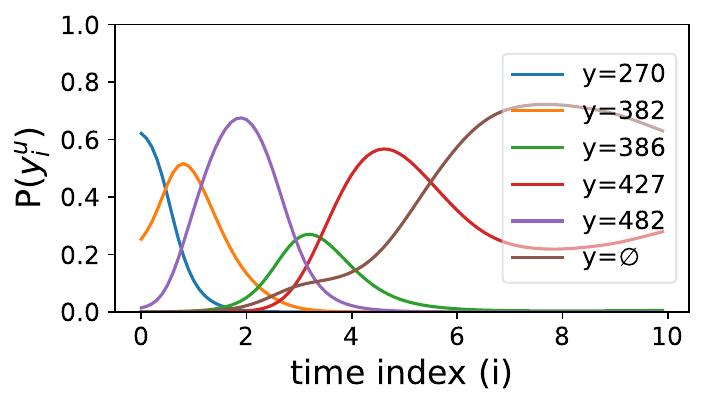}
        \captionsetup{skip=-2pt}
        \caption{P$(y_i^u)$: w/o padded training}
        \label{fig:c}
    \end{subfigure}
    % (d)
    \begin{subfigure}[b]{0.24\textwidth}
        \centering
        \includegraphics[width=\textwidth]{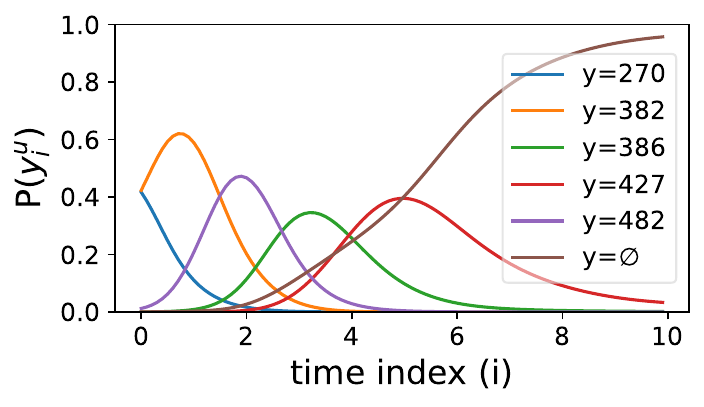}
        \captionsetup{skip=-2pt}
        \caption{P$(y_i^u)$: with padded training}
        \label{fig:d}
    \end{subfigure}
    
    \caption{Comparison of the adoption of padded training: (a) and (b) show the probability of $\varnothing$, while (c) and (d) show the probability of $y$ for a fixed $u$.}
    \label{fig2}
\end{figure}

We analyzed the training and inference schemes of SegINR. First, we compared the use of auxiliary padding during training with the $i_{pad}$ setting. The model shown in Fig.~\ref{fig2}(a) and Fig.~\ref{fig2}(c) was trained to predict $\varnothing$ only at $i=\textbf{d}_u+1$, while the model in Fig.~\ref{fig2}(b) and Fig.~2(d) was trained using auxiliary padded $\varnothing$. Additionally, we compared different inference schemes. In the first approach, $\varnothing$ is returned if its probability surpasses a threshold of $\tau=0.5$, whereas in the second approach, $\varnothing$ is returned when it has the highest probability among all candidates. The results are presented in Fig.~\ref{fig2} and Table~\ref{schemes}.

As shown in Fig.~\ref{fig2}, the model without padded training exhibited irregular probabilities for $\varnothing$ and $y$ due to the absence of extrapolated constraints. In contrast, padded training resulted in more consistent probabilities after $\varnothing$ reached its peak. Table~\ref{schemes} demonstrates that these tendencies significantly improve speech quality and intelligibility. The combination of padded training and the use of a threshold $\tau$ produced the highest quality, as it helps ensure the reliable duration modeling by stably returning of $\varnothing$.

\begin{table}
\caption{Character Error Rates~(CER) of each cases of using padded training }
\label{schemes}
\centering
% \begin{tabular}{@{}lc@{}c@{}rc@{}c@{}}
%\begin{tabular}{l r r r r}
\begin{tabular}{l c c c c}
\toprule
\textbf{CER}~(\%) & w/o padded training & with padded training\\ 
\midrule
infer w/o threshold $\tau$     & 8.55 & 7.66    \\
infer with threshold $\tau$    & 14.48 & \textbf{3.14}  \\
\bottomrule
\end{tabular}
\end{table}

\section{Conclusion}
We proposed SegINR, a novel framework for sequence alignment in TTS. By leveraging the concept of conditional INRs, we modeled frame-level speech features on a segment-wise basis and applied this approach to semantic token prediction tasks. Our results demonstrate the feasibility and superiority of SegINR. In future work, we plan to further explore the use of SegINR for generating other speech features and to integrate it with various generative models to enhance its generative capabilities.

\bibliographystyle{unsrt}
\bibliography{refs}

\end{document}